\documentclass[12pt]{article}
\usepackage{graphicx}

\newsavebox{\hflrar}
\sbox{\hflrar}{\makebox[0pt][l]
{${\scriptstyle \leftharpoonup}$}{${\scriptstyle \rightharpoonup}$}}

\def \to {\rightarrow}
\def \abst {\vert t \vert}
\def\bfsig{\mbox{\boldmath$\sigma$}}
\def\DT{\mbox{\boldmath$\Delta_T$}}

\def\bfej{\mbox{\boldmath$\varepsilon$}}

\begin{document}
\begin{flushright}
AS-ITP-2003-001
\end{flushright}
\pagestyle{plain}
\vskip 10mm
\begin{center}
{\Large\bf Diffractive Photoproduction of $\eta_c$} \\
\vskip 10mm
J. P. Ma   \\
{\small {\it Institute of Theoretical Physics , Academia
Sinica, Beijing 100080, China }} \\
~~~ \\
\end{center}

\vskip 0.4 cm


\begin{abstract}
Diffractive photoproduction of $\eta_c$ is an important process to
study the effect of Odderon, whose existence is still not confirmed
in experiment. A detailed interpretation of
Odderon in QCD, i.e., in terms of gluons is also unclear.
Taking charm quarks as heavy quarks, we can use NRQCD
and take $\eta_c$ as a $c\bar c$ bound state. Hence, in the production of
$\eta_c$
a free $c\bar c$ pair is first produced
and this pair is transformed into $\eta_c$ subsequently.
In the forward region of the kinematics,
the $c\bar c$ pair interacts with initial hadron through exchanges of
soft gluons. This interaction can be studied with HQET, which provides
a systematic expansion in the inverse of the $c$-quark mass $m_c$.
We find that the calculation of the $S$-matrix element in the
forward region can be formulated as the problem of solving a wave function
of a $c$-quark propagating in a background field of soft gluons.
At leading order
we find that the differential cross-section can be expressed with
four functions, which are defined with a twist-3 operator of gluons.
The effect of exchanging a Odderon can be identified with this operator
in our case.
We discuss our results in detail and compare them with those
obtained in previous studies. Our results and those from other studies
show that the differential cross-section is very small in the forward region.
We also show that the production through photon exchange
is dominant in the extremely forward region, hence
the effect of Odderon exchange can
not be identified in this region.
For completeness we also give results for diffractive
photoproduction of $J/\Psi$.

\end{abstract}

\vfill\eject\pagestyle{plain}\setcounter{page}{1}

\noindent
{\large\bf 1. Introduction}
\par\vskip20pt
Diffractive photoproduction of a pseudoscalar meson is an
interesting process because
the production is related to the postulated object: Odderon\cite{O1},
the partner of Pomeron.
The Pomeron is even under charge conjugation $C$, while
the Odderon has $C=-1$. The exchange of Pomeron and Odderon
is believed to delivery dominant contributions to hadronic
cross section at high energy. An interesting review about the history of Odderon
and relevant references can be found in \cite{Braun}.
Giving the importance of its existence in theory, experimentally
it is still unsuccessful to hunt the effect induced by Odderon. Only
one indication for existence of Odderon is found in the $t$-dependence
of $pp$- and $p\bar p$ elastic cross-section\cite{exp1}. Theoretically
the effect of Odderon exchange in $pp$- and $p\bar p$ scattering at high energy
has been intensively studied, a recent work and useful references
can be found in \cite{Dosch}.
\par
It has been suggested that the effect of Odderon may be detected
in diffractive pseudoscalar meson production for collision of a hadron with a real-
or virtual photon\cite{schmna,kina,Berger1,engel1,engel2,Bartels},
where effects of Pomeron are absent.
It is difficult to made predictions for the proposed processes
by starting from QCD directly. Some models are introduced to make
predictions for the $ep$ collider at HERA.
Unfortunately,
the experimental study in photoproduction of $\pi^0$\cite{exp2}
gives a negative result. In \cite{HPST} it is suggested
to look for Odderon in productions of two pions.
In \cite{engel1, engel2,Bartels}
diffractive photoproduction of $\eta_c$ are studied, predictions
may be made more reliably than those for production of light meson, because
of the following reasons: The structure
of $\eta_c$ is simpler than light mesons, the large quark mass $m_c$
enables us to use perturbative QCD at certain level and
the $c$-quark content in the initial hadron can be neglected.
Therefore the production of $\eta_c$ can be imagined as that the initial photon
is split into a $c\bar c$ pair, this pair exchanges gluons with the initial
hadron $h$ and forms the produced $\eta_c$ after exchanges. This is illustrated
in Fig.1. Effect of exchanging gluons can be thought as the effect
of exchanging a Odderon.
In \cite{engel1,engel2} one uses perturbative QCD
to handle the emission of gluons by the $c\bar c$ pair, at leading
order only three gluons are emitted. The interaction of the three
gluons with the initial hadron $h$ is described by a impact factor.
Using models for the impact factor one can obtain some numerical
predictions. In \cite{Bartels} one uses a set of new Odderon states
to account the effect of Odderon exchange and numerical predictions
can also be made. All of these works deliveries a total cross-section
for HERA roughly at order of $10^2$pb, but the $t$-dependence
is predicted differently in different works.
\par\vskip10pt

\begin{figure}[hbt]
\centering
\includegraphics[width=8cm]{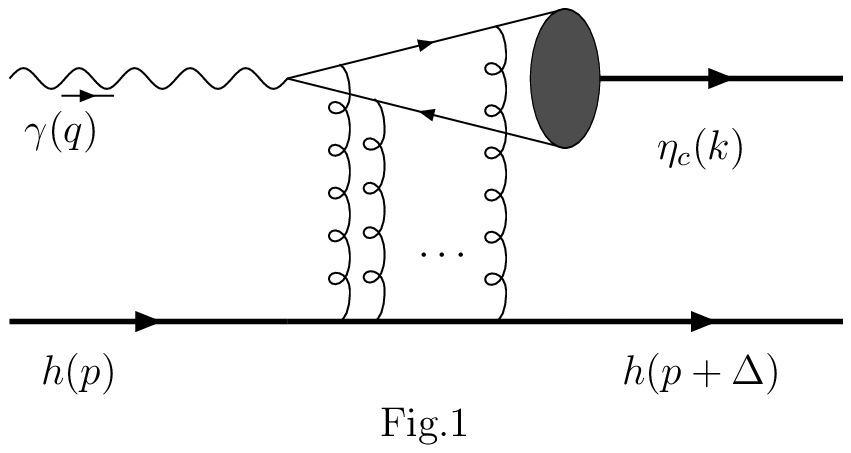}
\caption{Typical diagram for exchange of many soft gluons for
$\eta_c$ photoproduction }
\label{Feynman-dg1}
\end{figure}
\par\vskip 20pt

In this work we study the problem from another point of view.
We make an attempt to answer the question if one can express the differential cross-section
in the nearly forward direction with quantities which
are well defined in the framework of QCD. The answer is closely related
to a QCD interpretation of Odderon in our case.
If $\eta_c$ is produced diffractively and the beam energies are
large enough, the exchanged gluons are soft. It is questionable
to use perturbative QCD for these soft gluons. But, the charm quark
can be taken as a heavy quark, the emission of these soft gluons
can be studied with the Heavy Quark Effective Theory(HQET)\cite{HQET},
in which a systematical expansion in $m_c^{-1}$ can be made. Also
by taking charm quark as a heavy quark one can describe $\eta_c$
with nonrelativistic QCD(NRQCD)\cite{nrqcd}, in which
a systematic expansion in the small velocity $v_c$, which is the
velocity of a $c$- or $\bar c$ quark inside a charmonium in its rest frame,
can be made.
At leading order of $v_c$ the $c\bar c$ pair
after exchanging soft gluons can be taken as on-shell. Using this fact, the study of
the emission
of soft gluons by the $c$- or $\bar c$ quark can be formulated to solve
a wave function of the $c$- or $\bar c$ quark propagating in a background
field of soft gluons. The problem of light quarks propagating
in a background field of soft gluons have been studied in \cite{Na}
in relation to hadron-hadron scattering at high energy.
In our case these wave functions can be solved by expanding
them in $m_c^{-1}$.
It is interesting
to note that at the limit $m_c\to\infty$
the exchanged Odderon here consists of three soft gluons but in a special gauge.
\par
The approach  outlined above has been used to study similar cases with exchanges
of soft gluons like $J/\Psi\to\gamma^*+{\rm soft\ pions}$\cite{Ma,MaXue} and
diffractive photoproduction of $J/\Psi$\cite{MaXu}. In \cite{MaXu}
it is assumed that results for the $S$-matrix element can be obtained by taking two-gluon
exchange in a special gauge. In this work we derive these results
for completeness and without the assumption. It shows that for
the case of $J/\Psi$ the exchanging Pomeron consists of two gluons in
this special gauge.
\par
In this work we are unable to give numerical results in detail, because the differential
cross-section is expressed with four unknown functions, which
are defined with a twist-3 gluonic operator in QCD. However, order of magnitude
can be estimated in comparison with the case of $J/\Psi$. It turns out that
the differential cross-section is very small as that predicted by model
calculations\cite{engel1,engel2,Bartels}. With our result it can be show that
the differential cross-section can be non zero in the limit $t\to 0$, in contrast
to the predicted in \cite{engel1,engel2,Bartels}, where $t$ is the squared
momentum transfer between light hadrons. The reason for this
is that there are nonperturbative effects which can cause the helicity flip
of the initial hadron. This will be discussed in detail.
Since the differential cross-section with exchanges of soft gluons is small,
the exchange of
a photon, instead of exchanging many gluons in Fig.1., can have a sizeable
contribution. At first look, the amplitude with exchange of a photon
is divergent in the limit $t\to 0$. It should be noted that for large and finite $s$
the minimum of $\vert t\vert$ is small but not exactly zero. This fact makes
the amplitude finite, it leads to that the differential cross-section
in the forward direction increases linearly with $s$ and can be determined
completely.
We find that at energies relevant
to HERA the contribution from photon exchange is actually dominant in the
extremely forward direction.
This may exclude the possibility to observe effect of Odderon exchange
in this kinematical region. However, the contribution from photon exchange
decreases with increasing $\vert t\vert$ more rapidly than that of Odderon
exchange, it is possible to identify Odderon exchange at $\vert t\vert$
which is not very close to its minimal value.
\par
Our work is organized as the following: In Sect.2. we use NRQCD factorization
for $\eta_c$ and then formulate the problem of calculating
the $S$-matrix element
as the problem of solving  wave functions of $c$ quarks
in a background field of soft gluons. In Sect.3. we solve the wave functions
by expanding them in $m_c^{-1}$. The relevant part of solutions are given.
In Sect.4. we use these solutions to derive the $S$-matrix element for diffractive
photoproduction of $J/\Psi$. The results derived in this way are exactly the same
as those derived by the assumption of two-gluon exchange in a special gauge\cite{MaXu}.
In Sect.5. we derive the results for $\eta_c$ production and discuss our results.
In Sect.6. we study the contribution through exchange of a photon. Sect.7. is our
summary.

\par\vskip20pt
\noindent {\large\bf 2. Soft gluon exchange}
\par\vskip20pt
We consider the diffractive process:
\begin{equation}
\gamma (q) +h(p) \to h(p+\Delta) + \eta_c (k)
\end{equation}
where the momenta are indicated in the brackets and $k=q-\Delta$.
The Mandelstam
variables are defined as $
s=(q+p)^2$ and $t=\Delta^2$.
We consider the kinematic region where $\abst$ is at order
of $\Lambda_{QCD}^2$ and each component of $\Delta$ is at order
of $\Lambda_{QCD}$. The initial photon is real with $q^2=0$,
$h$ is any light hadron whose mass $m$ is at order
of  $\Lambda_{QCD}$. Throughout this work we take
nonrelativistic normalization for $\eta_c$ and $c$-quark.
The process undergoes like that the initial photon first splits into a
$c\bar c$ pair, this pair exchanges soft gluons with the light hadron
and forms $\eta_c$ after the exchanges. A typical diagram for this is given
in Fig.1.
The $S$-matrix element can be written as
a sum over contributions with $n$-gluon exchange:
\begin{eqnarray}
\langle f \vert S\vert i\rangle &=& (2\pi)^4 \delta^4(q-\Delta -k)\sum_n \frac{1}{n!}
\int d^4 x_1 d^4 x_2\cdots d^4 x_n d^4 x \int \frac {d^4 k_1}{(2\pi)^4}
\frac {d^4 k_2}{(2\pi)^4}\cdots \frac {d^4 k_n}{(2\pi)^4}
 \frac {d^4 q_1}{(2\pi)^4}
\nonumber\\
&& \cdot A_{ij}^{a_1a_2\cdots a_n, \mu_1\mu_2
\cdots \mu_n}(q,q_1,k_1, k_2,\cdots k_n) e^{-iq_1\cdot x} \langle \eta_c (k)\vert
\bar c_i(x) c_j(0) \vert 0\rangle
\nonumber \\
&& \cdot e^{ik_1\cdot x_1 +\cdots ik_n\cdot x_n}
\langle h(p+\Delta) \vert G_{\mu_1}^{a_1} (x_1) G_{\mu_2}^{a_2} (x_2)
\cdots G_{\mu_n}^{a_n} (x_n)\vert h(p)\rangle,
\end{eqnarray}
where $i$ and $j$ stand for color- and Dirac indices for the Dirac fields
$\bar c(x)$ and $c(0)$ respectively. $A_{ij}^{a_1a_2\cdots a_n, \mu_1\mu_2
\cdots \mu_n}(q,q_1,k_1, k_2,\cdots k_n)$ is the amplitude for that the initial photon
splits into a $c\bar c$ pair and this pair emits $n$ gluons. After the emission
the pair is in general not on-shell. In the rest frame of $\eta_c$ the $c$- or
$\bar c$-quark moves with a small velocity $v_c$. We can expand
the matrix element $\langle \eta_c (k)\vert
\bar c_i(x) c_j(0) \vert 0\rangle$ in the small $v_c$ with NRQCD fields. The
leading term in the expansion reads:
\begin{equation}
\langle \eta_c (k)\vert
\bar c_i(x) c_j(0) \vert 0\rangle =
-\frac{1}{6\sqrt{v^0}} \left [ \frac{1-\gamma\cdot v}{2}\gamma_5
 \frac{1+\gamma\cdot v}{2} \right ]_{ji} e^{im_cv\cdot x}
 \langle \eta_c \vert \psi^\dagger \chi \vert 0\rangle
  \cdot \left ( 1 +{\cal O} (v_c^2) \right ),
\end{equation}
where $v= k/M_{\eta_c}$ is the velocity of $\eta_c$, the matrix element is defined
with NRQCD fields in the rest frame of $\eta_c$, where $\psi^\dagger$ and $\chi$
are NRQCD fields of two components, $\psi^\dagger$ or $\chi$ creates
a heavy quark or a heavy antiquark respectively.
The matrix element can be determined
from the decay width for $\eta_c\to\gamma\gamma$:
\begin{equation}
\Gamma (\eta_c\to\gamma\gamma) =\frac {32\pi\alpha^2}{81 m_c^2}  \vert
  \langle \eta_c \vert \psi^\dagger \chi \vert 0\rangle \vert ^2.
\end{equation}
We will neglect higher orders of $v_c$ and only keep the leading term
in Eq.(3). In this approximation,  the $c$- and $\bar c$ quark after emission
of $n$ gluons is on-shell because of
the projection operators $(1\pm \gamma\cdot v)/2$. They carry the same momentum
$m_cv$ and the mass of $\eta_c$ is approximated by $M_{\eta_c}=2m_c$. Using
the spinors $\bar u (m_cv, s_1)$ and $v(m_cv, s_2)$ one can write the projection
operators as :
\begin{eqnarray}
 \frac{1+\gamma\cdot v}{2} &=& m_cv^0\sum_{s_1}u (m_cv, s_1)\bar u (m_cv, s_1),
 \nonumber\\
  \frac{1-\gamma\cdot v}{2} &=& -m_cv^0\sum_{s_2}v (m_cv, s_1)\bar v (m_cv, s_2),
\end{eqnarray}
where $s_1$ and $s_2$ represents the spin state of $c$- and $\bar c$ quark
respectively. Using Eq.(3) and Eq.(5) we obtain:
\begin{eqnarray}
\langle f \vert S\vert i\rangle &=& (2\pi)^4 \delta^4(q-\Delta -k)
\frac{1}{6\sqrt{v^0}} ( m_cv^0)^2
\langle \eta_c \vert \psi^\dagger \chi \vert 0\rangle
\sum_{s_1,s_2} \bar v (m_cv, s_2) \gamma_5 u (m_cv, s_1)
\nonumber\\
&& \sum_n \frac{1}{n!}
\int d^4 x_1 d^4 x_2\cdots d^4 x_n \int \frac {d^4 k_1}{(2\pi)^4}
\frac {d^4 k_2}{(2\pi)^4}\cdots \frac {d^4 k_n}{(2\pi)^4}
\nonumber\\
 && \cdot \bar u (m_cv, s_1) A^{a_1a_2\cdots a_n, \mu_1\mu_2
\cdots \mu_n}(q,m_cv,k_1, k_2,\cdots k_n) v (m_cv, s_1)
\nonumber \\
&& \cdot e^{ik_1\cdot x_1 +\cdots ik_n\cdot x_n}
\langle h(p+\Delta) \vert G_{\mu_1}^{a_1} (x_1) G_{\mu_2}^{a_2} (x_2)
\cdots G_{\mu_n}^{a_n} (x_n)\vert h(p)\rangle .
\end{eqnarray}
Clearly, the amplitude $\bar u (m_cv, s_1) A^{a_1a_2\cdots a_n, \mu_1\mu_2
\cdots \mu_n}(q,m_cv,k_1, k_2,\cdots k_n) v (m_cv, s_1)$ is just for
the process in which
the initial photon
splits into a $c\bar c$ pair, this pair emits $n$ gluons and becomes on-shell
when the emission is completed. Hence the $S$-matrix element can be written
as:
\begin{eqnarray}
\langle f \vert S\vert i\rangle &=& -i (2\pi)^4 \delta^4(q-\Delta -k)
\frac{1}{6\sqrt{v^0}} ( m_cv^0)^2
\langle \eta_c \vert \psi^\dagger \chi \vert 0\rangle
\sum_{s_1,s_2} \bar v (m_cv, s_2) \gamma_5 u (m_cv, s_1)
\nonumber\\
 && \cdot \langle h(p+\Delta), c(m_cv,s_1), \bar c(m_cv,s_2)\vert
    J_c^\mu (0) \vert h(p)\rangle \varepsilon_\mu,
\end{eqnarray}
where $\varepsilon$ is the polarization vector of the initial photon, $J_c^\mu$
is the charm quark part of the electro current, it is
$J_c^\mu = eQ_c \bar c\gamma^\mu c$ with $Q_c=2/3$. With the
approximation in Eq.(3) the production process
can be viewed as a two-step process, in which a on-shell $c\bar c$ pair is produced,
then this pair is converted into $\eta_c$. The probability amplitude for the conversion
is the matrix element $\langle \eta_c \vert \psi^\dagger \chi \vert 0\rangle$.
\par
With the $c$- and $\bar c$ quark in the final state one can apply the
standard LSZ reduction formula for the matrix element:
\begin{eqnarray}
&& \langle h(p+\Delta), c(m_cv,s_1), \bar c(m_cv,s_2)\vert
    \bar c (0) \gamma^\mu c(0) \vert h(p)\rangle    \nonumber\\
&& \  =
    \frac{1}{Z_2} \int d^4 x d^4 y e^{m_cv\cdot(x+y)}\bar u(m_cv, s_1)
    (i\gamma\cdot\partial_x -m_c)
    \nonumber\\
&& \ \ \ \cdot  \langle h(p+\Delta )\vert c(x)
    \bar c (0) \gamma^\mu c(0) \bar c(y) \vert h(p)\rangle
(-i \gamma\cdot \overleftarrow{\partial}_y -m_c) v(m_cv,s_2),
\end{eqnarray}
where $Z_2$ is the renormalization constant of $c$ quark field.
We can
evaluate the matrix element with help of the QCD path-integral, and perform
first the integration over c-quark field, while gluon fields and
other dynamical freedoms will be integrated later. After the
integration over c-quark field we obtain:
\begin{eqnarray}
&& \langle h(p+\Delta), c(m_cv,s_1), \bar c(m_cv,s_2)\vert
    \bar c (0) \gamma^\mu c(0) \vert h(p)\rangle    \nonumber\\
&& \  =-
    \frac{1}{Z_2} \int d^4 x d^4 y e^{m_cv\cdot(x+y)}\bar u(m_cv, s_1)
    (i\gamma\cdot\partial_x -m_c)
    \nonumber\\
&& \ \ \ \cdot  \langle h(p+\Delta) \vert S(x,0)\gamma^\mu S(0,y) \vert h(p)\rangle
(-i \gamma\cdot \overleftarrow{\partial}_y -m_c) v(m_cv,s_2),
\end{eqnarray}
where $S(x,y)$ is the $c$ quark propagator defined as
\begin{equation}
  S(x,y) =\frac{1}{i}\langle 0\vert c(x) \bar c(y) \vert 0\rangle.
\end{equation}
Because the gluon fields are not integrated at the moment and they can be
taken as a background. Then the propagator
describes how the quark propagates under the background of the gluon fields.
With the propagator we can introduce a wave function for the $c$- and $\bar c$ quark
respectively. We define:
\begin{eqnarray}
\bar\psi_c (x) &=& \bar u(m_cv,s_1)\int d^4 y e^{im_cv\cdot y}
(i\gamma\cdot\partial_y -m_c) S(y,x),
\nonumber\\
\psi_{\bar c} (x) &=& \int d^4y S(x,y)(-i \gamma\cdot \overleftarrow{\partial}_y -m_c)
  v(m_cv,s_2) e^{im_cv\cdot y},
\end{eqnarray}
with these wave functions we obtain for the matrix element:
\begin{equation}
\langle h(p+\Delta), c(m_cv,s_1), \bar c(m_cv,s_2)\vert
    \bar c (0) \gamma^\mu c(0) \vert h(p)\rangle
  =-\frac{1}{Z_2}
\langle h(p+\Delta) \vert \bar\psi_{c}(0)\gamma^\mu \psi_{\bar c}(0) \vert h(p)\rangle.
\end{equation}
It should be noted that the quark fields in above manipulations are bar fields,
especially, in Eq.(3). In Eq.(3) one should use the renormalized fields in both
sides to perform the expansion in $v_c$, it will give an extra factor of $Z_2$.
When we substitute Eq.(12) into Eq.(7), this extra factor $Z_2$ will cancel
the $Z_2$ in the denominator of Eq.(12). For gluon fields $G^\mu (x)$ they
always appear here as a product of $g_s G^\mu (x)$, which can be simply replaced
with renormalized quantities without introducing any extra factor.
With Eq.(12) we get the $S$-matrix element:
\begin{eqnarray}
\langle f \vert S\vert i\rangle &=& -ie (2\pi)^4 \delta^4(q-\Delta -k)
\frac{2}{18\sqrt{v^0}} ( m_cv^0)^2
\langle \eta_c \vert \psi^\dagger \chi \vert 0\rangle
\nonumber\\
 && \cdot
\sum_{s_1,s_2} \bar v (m_cv, s_2) \gamma_5 u (m_cv, s_1)
 \langle h(p+\Delta) \vert \bar\psi_{c}(0)\gamma\cdot\varepsilon\psi_{\bar c}(0) \vert h(p)\rangle.
\end{eqnarray}
\par
Similar results can also be obtained if we consider diffractive photoproduction
of $J/\Psi$:
\begin{equation}
\gamma (q) +h(p) \to h(p+\Delta) + J/\Psi (k).
\end{equation}
The expansion in the small $v_c$ corresponding to Eq.(3) reads:
\begin{eqnarray}
\langle J/\psi \vert\bar c_i(x) c_j(y) \vert 0\rangle
  &=& \frac{1}{6\sqrt{v^0}}e^{im_c v\cdot (x+y)}\left [\frac {(1-\gamma\cdot v )}{2}
    \gamma\cdot \varepsilon_J^*(v) \frac{(1+\gamma\cdot v)}{2}\right ]_{ji}
    \nonumber\\
    && \cdot \langle J/\psi \vert \psi^\dagger \bfsig\cdot\bfej_J ({\bf v}=0)
      \chi\vert 0\rangle  +\cdots ,
\end{eqnarray}
where $\varepsilon_J$ is the polarization vector of $J/\Psi$. The NQRCD matrix element
is related to the decay width of $J/\Psi\to\ell^+\ell^-$:
\begin{equation}
\Gamma (J/\psi\to e^+e^- ) = \alpha_{em}^2 Q_c^2 \frac {2\pi}{3m_c^2}
  \vert \langle J/\psi \vert \psi^\dagger\bfsig\cdot\bfej({\bf v}=0)
      \chi \vert 0\rangle \vert^2.
\end{equation}
With these one can obtain the $S$-matrix element for the case with $J/\Psi$:
\begin{eqnarray}
\langle f \vert S\vert i\rangle &=& ie (2\pi)^4 \delta^4(q-\Delta -k)
\frac{2}{18\sqrt{v^0}} ( m_cv^0)^2
\langle J/\psi \vert \psi^\dagger\bfsig\cdot\bfej({\bf v}=0)
      \chi \vert 0\rangle
\nonumber\\
 && \cdot
\sum_{s_1,s_2} \bar v (m_cv, s_2) \gamma\cdot \varepsilon_J^*(v) u (m_cv, s_1)
 \langle h(p+\Delta) \vert \bar\psi_{c}(0)
 \gamma\cdot\varepsilon\psi_{\bar c}(0) \vert h(p)\rangle.
\end{eqnarray}
\par
The wave functions defined in Eq.(11) satisfy the Dirac equations:
\begin{eqnarray}
\bar\psi_c(x)\left ( i \gamma\cdot \overleftarrow{D} +m_c\right ) &=& 0,
\nonumber\\
\left ( i \gamma\cdot D -m_c\right )\psi_{\bar c} (x) &=& 0
\end{eqnarray}
with the boundary conditions:
\begin{eqnarray}
\bar\psi_c(x) &\to & \bar u(m_cv,s_1)e^{im_cv\cdot x},
\nonumber\\
\psi_{\bar c} (x) &\to & v(m_cv,s_2)e^{im_cv\cdot x}
\end{eqnarray}
for the time  $x^0\to\infty$. $D^\mu =\partial^\mu + ig_s G^\mu(x)$ is the covariant derivative.
These wave functions
describe the behavior of $c$- or $\bar c$ quark moving under a background
of gluon fields. As discussed before, the exchanged gluons between the $c\bar c$
pair and the light hadron $h$ are soft, this results in that the background gluon
fields are mainly with long-wave lengthes and have a weak dependence on $x$,
the dependence is characterized by the scale $\Lambda_{QCD}$. In the heavy quark
limit, $m_c\gg \Lambda_{QCD}$, this enables us to solve the equations by expanding
the wave functions in $m_c^{-1}$. Using the solutions we can obtain the matrix element
in Eq.(12) in terms of gluon fields. Then we integrate over the gluon fields, i.e.,
complete the QCD path integral mentioned after Eq.(8), and finally we
can get the $S$-matrix element for the diffractive process. It should be noted
that the propagator in Eq.(10) is a Feynman
propagator. The wave-functions defined in Eq.(11) do satisfy the Dirac equation
with a background field of gluons, but they do not have a simple boundary
condition like in Eq.(19). However, it was shown \cite{Na} that the Feynman
propagator in Eq.(10) can be replaced with a advanced- or retarded propagator,
if the
background field varies enough slowly with the space-time. Therefore one can have
solutions of Eq.(18) and they satisfy the boundary conditions in Eq.(19). In
our case we will make the expansion in $m_c^{-1}$ and in this expansion the
c-quark and $\bar{c}$ quark are decoupled, and it results in that the
Feynman propagator will automatically be an advanced propagator for $c$ and
$\bar{c}$ in our case.
\par
\vskip20pt
\noindent
{\large\bf 3. The wave functions in the expansion in $m_c^{-1}$}
\par\vskip20pt
In this section we will solve the Dirac equations up to order of $m_c^{-2}$
for our purpose. With the invention of HQET the expansion in $m_c^{-1}$
is now quite standard. Taking $\psi_{\bar c}$ as an example, we can decompose
the wave function as:
\begin{equation}
\psi_{\bar c} (x) = e^{im_c v\cdot x} \left ( h_+(x) + h_-(x)\right ),
\end{equation}
where
\begin{equation}
\gamma\cdot v h_+(x) = h_+(x), \ \ \ \  \gamma\cdot v h_-(x) =-h_-(x).
\end{equation}
To present the solution for $\psi_{\bar c}$ we introduce some notations. For any
vector $A$ one can decompose it as:
\begin{equation}
A^\mu = v\cdot A v^\mu + A^\mu_T,\ \ \ \ \ {\rm with}\ v\cdot A_T =0.
\end{equation}
Using the Dirac equation for $\psi_{\bar c}$ one can express
$h_+(x)$ in term of $h_-(x)$ as a series of $m_c^{-n}$:
\begin{eqnarray}
h_+(x) &=& \frac {1}{2m_c-iv\cdot D} i\gamma\cdot D_T h_-(x)
\nonumber\\
      &=& \left (\frac{1}{2m_c}+ \frac{iv\cdot D}{4m_c^2} \right )
        i\gamma\cdot D_T h_-(x) +{\cal O}(\frac{1}{m_c^3})
\end{eqnarray}
while $h_-(x)$ is constrained by the equation:
\begin{equation}
\left \{ iv\cdot D -i\gamma\cdot D_T \left (
\frac{i\gamma\cdot D_T}{2m_c}+ \frac{iv\cdot D i\gamma\cdot D_T}
 {4m_c^2} \right )\right \}
 h_-(x) =0 +{\cal O}(\frac{1}{m_c^3}).
\end{equation}
This equation can be solved by expanding $h_-(x)$ in $m_c^{-1}$, and eventually
we obtain the wave function. It is straightforward to solve this equation.
For convenience we define a gauge link $V(x)$ with $x^\mu=\omega v^\mu +x_T^\mu$
as:
\begin{equation}
V(x) = P \exp \left [ -ig_s \int_\omega^\infty d\tau v\cdot G(\tau, x_T) \right ]
\end{equation}
where we denote the $x$-dependence of $G^\mu(x)$ as $G^\mu(\omega,x_T)$
and $P$ stands for path ordering.
The leading order results for wave functions read:
\begin{eqnarray}
\psi_{\bar c} (x) &=& e^{im_cv\cdot x} V^\dagger (x) v(m_cv,s_2) +
{\cal O}(\frac{1}{m_c}),
\nonumber\\
\bar\psi_c (x) &=& e^{im_cv\cdot x} \bar u(m_cv,s_1) V(x)+
{\cal O}(\frac{1}{m_c}).
\end{eqnarray}
With these wave functions at the order of $m_c^0$ it is easy to find out
that the $S$-matrix element in Eq.(13) or Eq.(17) is zero because
$VV^\dagger =1$. To go beyond the leading order we make a gauge transformation:
\begin{eqnarray}
\psi_{\bar c} (x) &\to & V(x) \psi_{\bar c} (x),
\nonumber\\
\bar\psi_c (x)  &\to & V^\dagger (x) \bar\psi_c (x),
\nonumber \\
   G^\mu (x) &\to & V(x) G^\mu(x) V^\dagger(x) -\frac{i}{g_s}
     V(x) \partial^\mu V^\dagger (x).
\end{eqnarray}
The $S$- matrix element is invariant under the gauge transformation.
The transformed gauge field has no component along the direction $v$, i.e.,
$v\cdot G(x)=0$.
This is equivalent by taking the gauge $v\cdot G(x) =0$. To avoid introduction
of too many notations we will use the same notations for transformed fields. The
fields below should be understood as transformed fields or fields with the gauge
$v\cdot G (x) =0$. Now for gluon fields we have:
\begin{equation}
   v\cdot G(x) =0,\ \ \ \ \  v_\mu G^{\mu\nu}(x) =  v_\mu \partial^\mu G^\nu (x).
\end{equation}
\par
At the order of $m_c^{-1}$ the solutions for the wave functions
read:
\begin{eqnarray}
\psi_{\bar c} (x) &=& e^{im_cv\cdot x} \left \{ 1 + \frac{i}{2m_c}
    \gamma\cdot D_T -
   \frac {i}{2m_c} \int^\infty_\omega d\tau (\gamma\cdot D_T)^2 \right\}
   v(m_cv,s_2) +
{\cal O}(\frac{1}{m_c^2})
\nonumber\\
  &=& e^{im_cv\cdot x} \left \{ 1 +
   \frac {ig_s^2}{2m_c} \int^\infty_\omega d\tau G(\tau,x_T)\cdot G(\tau,x_T)
     +\cdots \right\}
   v(m_cv,s_2) +{\cal O}(\frac{1}{m_c^2})
\nonumber\\
\bar\psi_c (x) &=& e^{im_cv\cdot x} \bar u(m_cv,s_1) \left \{ 1
        -\frac{i}{2m_c}
    \gamma\cdot \overleftarrow{D}_T -
   \frac {i}{2m_c} \int^\infty_\omega d\tau (\gamma\cdot\overleftarrow {D}_T)^2 \right\}
              +{\cal O}(\frac{1}{m_c^2}) \nonumber\\
     &=&  e^{im_cv\cdot x} \bar u(m_cv,s_1) \left \{ 1 +
   \frac {ig_s^2}{2m_c} \int^\infty_\omega d\tau G(\tau,x_T)\cdot G(\tau,x_T)
     +\cdots \right\} +{\cal O}(\frac{1}{m_c^2}),
\end{eqnarray}
where in the $\{ \cdots\}$'s with $\cdots$ we have detailed terms,
which will lead to contributions to the $S$-matrix element in the case
with $J/\Psi$, and $\cdots$ denotes irrelevant terms. At this order
the $S$-matrix element in the case with $\eta_c$ is zero. To obtain
the nonzero $S$-matrix element one needs the solutions at order
of $m_c^{-2}$.
\par
At order of $m_c^{-2}$ there are many terms for the wave function
$\psi_{\bar c}(x)$. However, only one term will lead to a nonzero
contribution to the $S$-matrix element with $\eta_c$. We will only
give this term in detail for the solutions, other irrelevant terms
are denoted by $\cdots$. For $\psi_{\bar c} (x)$ we have:
\begin{eqnarray}
\psi_{\bar c} (x) &=& e^{im_cv\cdot x} \left \{ 1 + \frac{1}{m_c}
     \left \{ \cdots  \right \}
     -\frac{1}{4m_c^2} \left (\int^\infty_\omega d\tau_1 (\gamma\cdot D_T)^2
        \int^\infty_{\tau_1} d\tau_2 (\gamma\cdot D_T)^2  +\cdots\right )
        \right\}v(m_cv,s_2)
        \nonumber\\
        && +{\cal O}(\frac{1}{m_c^3}),
\end{eqnarray}
one can use the identity
\begin{equation}
  \left ( \gamma\cdot D \right )^2
     = D\cdot D +\frac{g_s}{2} \sigma_{\mu\nu} G^{\mu\nu}
\end{equation}
to indicate the relevant term more clearly. The part of wave functions
relevant to the case with $\eta_c$ reads:
\begin{eqnarray}
\psi_{\bar c} (x) &=& e^{im_cv\cdot x}  \{ 1 + \frac{1}{m_c}
      \{ \cdots   \}
     -\frac{g_s^2}{4m_c^2}  [\int^\infty_\omega d\tau_1 \int^\infty_{\tau_1} d\tau_2
        (g_s G\cdot G -\frac{1}{2} \sigma_{\mu\nu}
          G^{\mu\nu} )(\tau_1, x_T)
          \nonumber\\
    && \cdot  (g_s G\cdot G -\frac{1}{2} \sigma_{\mu\nu}
          G^{\mu\nu} )(\tau_2, x_T)  +\cdots ]
        \}v(m_cv,s_2)
         +{\cal O}(\frac{1}{m_c^3}),
  \nonumber\\
  \bar\psi_c (x) &=& e^{im_cv\cdot x} \bar u(m_cv,s_1) \{ 1 + \frac{1}{m_c}
      \{ \cdots   \}
     -\frac{g_s^2}{4m_c^2}  [\int^\infty_\omega d\tau_1\int^\infty_{\tau_1} d\tau_2
        (g_s G\cdot G -\frac{1}{2} \sigma_{\mu\nu}
          G^{\mu\nu} )(\tau_2, x_T)
          \nonumber\\
    && \cdot (g_s G\cdot G -\frac{1}{2} \sigma_{\mu\nu}
          G^{\mu\nu} )(\tau_1, x_T)  +\cdots ]
        \}
         +{\cal O}(\frac{1}{m_c^3}).
\end{eqnarray}
With the terms for the solutions, given in detail in Eq.(29) and Eq.(32),
we can evaluate the $S$-matrix element for $J/\Psi$ and $\eta_c$.
\par\vskip20pt
\noindent
{\large\bf 4. The $S$-matrix element for $J/\Psi$}
\par\vskip20pt
In this section we use the solutions of wave functions to derive the $S$-matrix
element for diffractive photoproduction of $J/\Psi$. Using Eq.(29) we obtain
the term in Eq.(17)
\begin{eqnarray}
 && (m_cv^0)^2\sum_{s_1,s_2} \bar v (m_cv, s_2) \gamma\cdot \varepsilon_J^*(v) u (m_cv, s_1)
 \langle h(p+\Delta) \vert \bar\psi_{c}(0)
 \gamma\cdot\varepsilon\psi_{\bar c}(0) \vert h(p)\rangle
 \nonumber\\
 && = \frac{ig_s^2}{m_c} \varepsilon^*_J \cdot \varepsilon
  \int^\infty_0 d\omega \langle h(p+\Delta) \vert G^{a,\mu} (\omega,0)
   G_\mu ^a (\omega,0)\vert h(p)\rangle.
\end{eqnarray}
Using translation covariance and
\begin{equation}
\int^\infty_0 d\omega e^{i\omega v\cdot\Delta} =\frac{i}{v\cdot\Delta +i0^+}
\end{equation}
where $0^+$ is a positive infinitesimal number, we obtain
\begin{eqnarray}
 && (m_cv^0)^2\sum_{s_1,s_2} \bar v (m_cv, s_2) \gamma\cdot \varepsilon_J^*(v) u (m_cv, s_1)
 \langle h(p+\Delta) \vert \bar\psi_{c}(0)
 \gamma\cdot\varepsilon\psi_{\bar c}(0) \vert h(p)\rangle
 \nonumber\\
 && = -\frac{g_s^2}{m_c^2} \varepsilon^*_J \cdot \varepsilon
   \langle h(p+\Delta) \vert G^{a,\mu} (0,0)
   G_{\mu} ^a (0,0)\vert h(p)\rangle.
\end{eqnarray}
In the above equation we has used the kinematics for the diffractive process:
\begin{equation}
v\cdot\Delta = -m_c \left ( 1+ \frac{t}{4m_c^2}\right )\approx -m_c.
\end{equation}
\par
For the gauge field $G^\mu$ with $v\cdot G =0$ as in Eq.(28) one can relate it
to the gluon field strength:
\begin{equation}
G^\mu (\omega, x_T) = -\int^\infty_{-\infty} d\tau \theta(\tau-\omega)
   v_\nu G^{\nu\mu}(\tau, x_T),
\end{equation}
where the step function is defined as:
\begin{equation}
\theta (t)=\int_{-\infty }^\infty \frac{d\omega }{2\pi }\frac{ie^{-i\omega t}%
}{\omega +i0^{+}}=\left\{
\begin{array}{ll}
1,  t\ge 0 \\
0,  t<0
\end{array}
\right.
\end{equation}
With this relation one can show that
\begin{eqnarray}
&& \langle h(p+\Delta) \vert G_T^{a,\mu}(0,0)
   G_{T\mu} ^a (0,0)\vert h(p)\rangle
   = \frac{4}{m_c} \int d z \frac{1}{1+z-i0^+}\cdot
         \frac{1}{1-z-i0^+} \nonumber\\
    && \cdot \int \frac{d\tau}{2\pi} g_s^2 e^{im_c z\tau}
         v_\mu v_\nu   \langle h(p+\Delta) \vert G^{a,\mu\rho} (\tau ,0)
   G^{a,\nu}_{\ \ \ \rho}(-\tau ,0)\vert h(p)\rangle.
\end{eqnarray}
Finally we obtain the $S$-matrix element for $J/\Psi$:
\begin{eqnarray}
\langle f \vert S\vert i\rangle &=& ie (2\pi)^4 \delta^4(q-\Delta -k)
\frac{4}{9\sqrt{v^0}}
\langle J/\psi \vert \psi^\dagger\bfsig\cdot\bfej({\bf v}=0)
      \chi \vert 0\rangle \frac{\varepsilon_J^*\cdot \varepsilon}{m_c^3}
\nonumber\\
 && \cdot\int d z \frac{1}{1+z-i0^+}\cdot
 \frac{1}{1-z-i0^+} \cdot F_R(z),
\end{eqnarray}
with
\begin{equation}
F_R(z) = \int \frac{d\tau}{2\pi} g_s^2 e^{im_c z\tau}
         v_\mu v_\nu   \langle h(p+\Delta) \vert G^{a,\mu\rho} (\tau v)
   G^{a,\nu}_{\ \ \ \rho}(-\tau v)\vert h(p)\rangle.
\end{equation}
This result is exactly the same as that given in \cite{MaXu},
where we use perturbative QCD
by taking only two gluon-exchange in the gauge $v\cdot G=0$.
Here we derive the same result without
using perturbative QCD, instead we only use the expansion in $m_c^{-1}$ in
an arbitrary gauge. The above results are expressed with the transformed gauge
field, the gauge transformation can be found in Eq.(27).  If we express
the results with the untransformed gauge field,
a gauge link will appear automatically
between the two operators of gluon strength fields.
The gauge link is along the direction
of $v$ with the gauge fields in the adjoint representation,
it starts at $x=-\tau v$ and ends at $x=\tau v$.
With the gauge link the results are gauge-invariant. It is
interesting to note that models with two-gluon exchange are widely used
for diffractive photoproduction of vector meson. Our result here shows that
such a model is a correct approximation in the heavy quark limit.
\par
In our approach, although a formal expansion in $m_c^{-1}$ is employed,
but the true expansion parameter is $(m_c v\cdot k)^{-1}$, where $k$ is
the momentum of a exchanged gluon, this can be
realized by inspecting the lagrangian of HQET and it is discussed
in detail in \cite{MaXu}. In this expansion, transversal momenta of exchanged gluons
are neglected at the leading order. Hence
the exchanged gluons will not resolve the structure of a heavy quarkonium
in the directions transversal to $v$. The situation here is similar to
the multi-pole expansion for gluon fields in hadronic transitions
like $\psi^\prime\to J/\Psi +\pi +\pi$ in \cite{MPE}.
\par
For sufficiently large beam energies, the dominant contribution to the
correlation function $F_R(z)$ is from the standard gluon-operators with
twist 2 and these operators are used to define the gluon distribution
of $h$. If we only take the dominant contribution, then the $S$-matrix
element is related to the generalized gluon distribution. However, one
can show that the $S$-matrix element can be related to the usual
gluon distribution $g_h(x)$ at $x=2m_c^2/s$ for $t\to 0$, this is different
than that in previous approaches\cite{THJ},
details can be found in \cite{MaXu}.
\par
In \cite{MaXu} it shows that the predicted cross-section has large deviation
from experimentally measured for $J/\Psi$, while for $\Upsilon$
the agreement is fairly good. One of possible reasons for the
large deviation can be corrections from higher orders in $m_c^{-1}$,
because $m_c$ may not be large enough, while for $\Upsilon$
the $b$-quark mass is heavy enough to have reliable results
from leading order.
At leading order the exchange is of two gluons as indicated
in Eq.(29). It is interesting
to note that with the method developed here, it is possible
to resume many exchanges of this type of two gluons, the resummation
will reduce the corrections from higher order in $m_c^{-1}$. Works
along this direction are under progress.

\par\vskip20pt
\noindent
{\large \bf 5. The $S$-matrix element for $\eta_c$}
\par\vskip20pt
Using the wave functions given in Eq.(30), it is straightforward to obtain the
$S$-matrix element for $\eta_c$. We obtain the corresponding part given
in Eq.(33):
\begin{eqnarray}
&& (m_cv^0)^2\sum_{s_1,s_2} \bar v (m_cv, s_2) \gamma_5 u (m_cv, s_1)
 \langle h(p+\Delta) \vert \bar\psi_{c}(0)
 \gamma\cdot\varepsilon\psi_{\bar c}(0) \vert h(p)\rangle
 \nonumber\\
 &&
   =-\frac{g_s^3}{8m_c^2} \varepsilon_{\sigma\rho\mu\nu} \varepsilon^\sigma
   v^\rho
  \int^\infty_{-\infty} d\tau_1 \int_{-\infty}^\infty d\tau_2\theta(\tau_1)
   \theta(\tau_2-\tau_1)
   \nonumber\\
  && \cdot  d^{abc} \langle h(p+\Delta) \vert \left\{ G^{a,\beta} (\tau_1v)
   G_\beta ^b (\tau_1v) G^{c,\mu\nu}(\tau_2v) +
   (\tau_1\to\tau_2,\tau_2\to\tau_1)
     \right\}
   \vert h(p)\rangle.
\end{eqnarray}
With help of Eq.(37) one can transform the gluon fields $G^\mu$ into the field
strength $G^{\mu\nu}$. After some algebra manipulation we can express the result
as:
\begin{eqnarray}
\langle f \vert S\vert i\rangle &=& ie (2\pi)^4 \delta^4(q-\Delta -k)
\frac{g_s^3}{72m_c^3\sqrt{v^0}}
\langle \eta_c \vert \psi^\dagger \chi \vert 0\rangle
  \varepsilon_{\sigma\rho\mu\nu} \varepsilon^\sigma
   v^\rho
  \int d\omega_1 d\omega_2 \cdot {\cal F}^{\mu\nu} (\omega_1,\omega_2)
   \nonumber\\
  && \cdot\left\{ \frac{1}{\omega_1+i0^+}\cdot \frac{1}{\omega_2+i0^+}\right\}
  \cdot\left\{ \frac{1}{m_c+\omega_1 +\omega_2 -i0^+} - \frac{1}{\omega_1+\omega_2 +i0^+}
  \right\},
\end{eqnarray}
with
\begin{equation}
{\cal F}^{\mu\nu}(\omega_1,\omega_2) = \int \frac{d\tau_1}{2\pi}
   \frac{d\tau_2}{2\pi} e^{-i\omega_1\tau_1-i\omega_2\tau_2}
 \cdot  d^{abc} \langle h(p+\Delta) \vert  G^{a,\sigma\alpha} (\tau_1v)
   G^{b,\rho}_{\ \ \alpha} (\tau_2v) G^{c,\mu\nu}(0)
   \vert h(p)\rangle v_\sigma v_\rho.
\end{equation}
Again, in the definition of ${\cal F}^{\mu\nu}(\omega_1,\omega_2)$ the
gluon fields are transformed fields in Eq.(27) or those in the gauge
$v\cdot G(x)=0$. If we use the untransformed to express Eq. (44),
the gauge links discussed in the last section will automatically
appear between the field strength operators.
Therefore, the $S$-matrix is gauge invariant. From this result one can also see
that in the heavy quark limit, the exchange of three gluons in the gauge
$v\cdot G(x)=0$ is responsible for the process and the polarizations of
the three gluons are all transversal to the direction $v$, and these three gluons
are emitted by $c$- or $\bar c$ quark. In other gauge, because
the appearance of the above mentioned gauge links, there are additionally
exchanges of gluons with infinite numbers, those gluons have polarizations
proportional to the direction of $v$ and their effects are included in the
gauge link.
\par
If the beam energies are large, which is really required by that $t \sim
\Lambda_{QCD}^2$, the vector $v$ approaches to a light-cone vector,
the dominant contribution for ${\cal F}^{\mu\nu}$ can be identified.
For convenience we take a coordinate system in which the photon moves
in the $-z$-direction and the hadron $h$ in the $z$-direction. We
introduce a light-cone coordinate system, components of a vector $A$ in this
coordinate system are related to those in the usual coordinate system as
\begin{equation}
 A^\mu =\left( A^+,A^-,{\bf A_T}\right) =
       \left( \frac{A^0+A^3}{\sqrt{2}},\frac{A^0-A^3}{\sqrt{2}},A^1,A^2\right).
\end{equation}
We introduce two light-cone vectors $n=(0,1,0,0)$ and $l=(1,0,0,0)$ with $l\cdot n =1$.
The momenta in the process  can be approximated in the limit
$s\approx 2k^-p^+\to\infty$ as:
\begin{eqnarray}
k^\mu &=& (0,k^-,{\bf 0_T})=k^-n^\mu, \nonumber\\
p^\mu &=& (p^+,\frac{m^2}{2 p^+},{\bf 0_T}) \approx
        p^+ l ^\mu, \nonumber\\
\Delta^\mu &=& \left( \frac{t-M_{J/\psi}^2}{2k^-},
                   \frac {-t-\frac{m^2}{p^+}\Delta^+} {2p^+},
                   \DT\right) \approx \left( -\frac{2m_c^2}{k^-},
                   -\frac{t}{2p^+},\DT\right)
       \nonumber\\
       v^\mu &=& \frac{(k-\Delta)^\mu}{M_{J/\psi}}\approx \left( \frac{m_c}{k^-},
       \frac{k^-}{2m_c},{\bf 0_T} \right)\approx
       \frac{k^-}{2m_c}n^\mu,
\end{eqnarray}
with the above approximated momenta and re-arrangement of variables
the dominant contribution of the $S$-matrix
element reads:
\begin{eqnarray}
\langle f \vert S\vert i\rangle & \approx & ie (2\pi)^4 \delta^4(q-\Delta -k)
\frac{g_s^3}{36m_c^3  \sqrt{v^0}}
\langle \eta_c \vert \psi^\dagger \chi \vert 0\rangle
  \varepsilon_{\sigma\rho\mu\nu} \varepsilon^\sigma n^\rho l^\mu
  \int d\omega_1 d\omega_2 \cdot {\cal W}^{\nu} (\omega_1,\omega_2)
   \nonumber\\
  && \cdot\left\{ \frac{1}{\omega_1+i0^+}\cdot \frac{1}{\omega_2+i0^+}\right\}
  \cdot\left\{ \frac{1}{2x_c+\omega_1 +\omega_2 -i0^+} - \frac{1}{\omega_1+\omega_2 +i0^+}
  \right\},
\end{eqnarray}
with $x_c=2m_c^2/s$ and
\begin{equation}
{\cal W}^{\nu}(\omega_1,\omega_2) =\frac{1}{p^+} \int \frac{d\tau_1}{2\pi}
   \frac{d\tau_2}{2\pi} e^{-i\omega_1\tau_1p^+-i\omega_2\tau_2p^+}
 \cdot  d^{abc} \langle h(p+\Delta) \vert  G^{a,+\alpha} (\tau_1n)
   G^{b,+}_{\ \ \ \alpha} (\tau_2n) G^{c,+\nu}(0)
   \vert h(p)\rangle,
\end{equation}
where $\omega_1$ and $\omega_2$ is dimensionless and $\nu$ can only be $1$ or $2$.
The function ${\cal W}^{\nu}(\omega_1,\omega_2)$ with $\nu=1,2$ is invariant under
a Lorentz boost along the $z$-direction. ${\cal W}^{\nu}$ is defined
by the twist-3 operator in the matrix element and its dimension
is 1 in mass, hence it is proportional to $\Lambda$, where $\Lambda$
can be one of the small scales like $\Lambda_{QCD}$, $\sqrt{-t}$, etc..
This is our main result for any light hadron. If one takes the approach
with exchange of a Odderon for the process, the effect of the Odderon
is then represented by ${\cal W}^i$, which is defined in the framework
of QCD.
\par
Now we consider the case that $h$ is a proton, for this case ${\cal W}^i(i=1,2)$
can be parameterized with four functions in general:
\begin{equation}
{\cal W}^i = \frac{1}{p^+} \bar u(p+\Delta )
\left[ f_1 i\sigma^{+i} +f_2 \frac{p^+\Delta^i}{m^2}
  +f_3 \frac{\Delta^i \gamma^+}{m}
  +f_4 \frac{p^+\gamma^i}{m} \right ] u(p),
\end{equation}
where $u$ and $\bar u$ is the spinor of the proton, $m$ is the proton mass. A similar
decomposition of quark operator can be found in \cite{Diehl,HoodJi}, a general
discussion about how to write these form factors in similar cases
is given in \cite{Diehl}.
The functions
$f_i(i=1,2,3,4)$ are with dimension 1 in mass and are with variables: $\omega_1$,
$\omega_2$, $t$ and $\xi =-\Delta^+/(2p^++\Delta^+)$. These functions
are proportional to $\Lambda$. The differential cross section
can be expressed with these four functions, the expression is too long to present here.
However one can look at how the differential cross-section depends on
dimensional quantities, like $s$, $t$, $m_c$ etc. . We have for
$t\to 0$:
\begin{equation}
\frac{d\sigma}{dt}(\eta_c) \sim \alpha_s^3 \frac
{\vert \langle \eta_c \vert \psi^\dagger \chi \vert 0\rangle\vert^2}{m_c^3}
\cdot \frac{\Lambda^2}{m_c^2}\cdot \frac{1}{s^2} \hat f(s,t,m_c^2),
\end{equation}
the dimensionless function $\hat f$ is determined by the four functions given
above. It is interesting to note that in the limit $\Delta^\mu \to 0$
$d\sigma/dt$ does not approach to zero. In this limit the contributions
from $f_2$ and $f_3$ indeed vanish , but the contributions from
$f_1$ and $f_4$ survive in the limit. This fact is in contrast
to the results from \cite{engel1,engel2,Bartels}. This result
can be understand as the following: Because the initial photon
has the helicity $\lambda_\gamma =\pm 1$ and $\eta_c$ has
the helicity 0, the proton will change its $J_z$ with $\pm 1$, the
component in $z$-direction of the total angular momentum.
This change can be made with change of $l_z$, the component
of the orbital angular momentum, whose effects are parameterized
by $f_2$ and $f_3$ and the change is represented by $\Delta^i$ in
Eq.(49). It is clear that in the limit $\Delta\to 0$ the contributions
from $f_2$ and $f_3$ go to zero. However, this change can be made
by changing the helicity of the proton, which can be identified
in massless limit as $s_z$,
the component in $z$-direction of the proton spin,  effect due to this is given
by $f_1$, the term with $f_4$ conserves helicity, but $s_z$ can not be exactly
identified with the finite mass $m$ as helicity, if $m$ is exactly zero,
the term does not exist in the limit $\Delta^\mu =0$, hence the nonzero
contribution from $f_4$ in the limit $\Delta^\mu =0$ is due to the finite mass
$m$. These contributions survive in the limit and there
are in general no reasons why they should vanish.
If one uses
perturbative QCD and takes a set of light-cone wave functions
of the proton to calculate ${\cal W}^i$ or the impact factor
in \cite{engel1,engel2} for interactions between three gluons
and the proton, where one takes a proton as a bound state of the three
quarks $uud$
and the three quarks are in $s$-wave,
then one would find that $f_1=0$
because perturbative QCD conserves helicities.
However, if one realizes that fact that the three quarks can have
orbital motions, whose effects are parameterized with wave-functions given
in \cite{JMY}, then $f_1$ is not zero. In general perturbative QCD
is questionable for this calculation, because
the gluons are soft. However, this indicates $f_1\neq 0$.
\par
\par\vskip20pt

\begin{figure}[hbt]
\centering
\includegraphics[width=8cm]{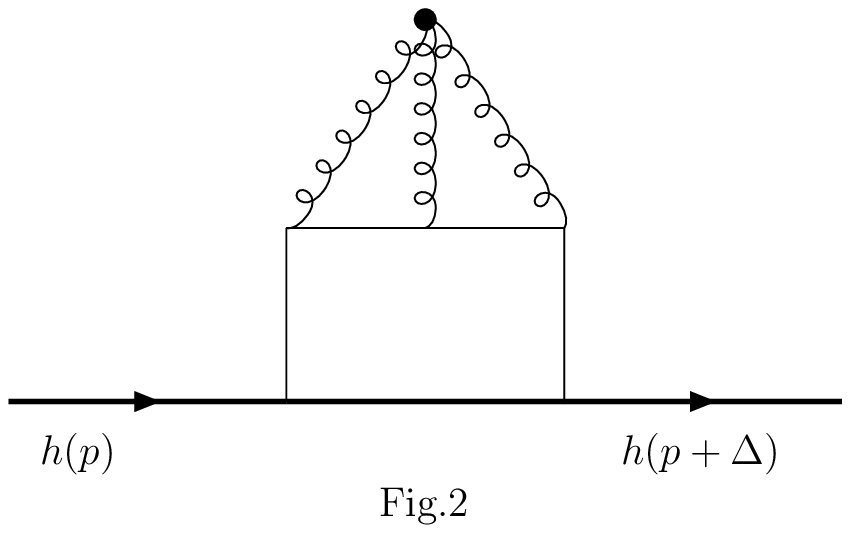}
\caption{Gluons are generated by a single quark in a proton, see text}
\label{Feynman-dg2}
\end{figure}
\par\vskip 20pt
If one thinks that the gluons, whose effect is represented by ${\cal W}^i$,
may be generated by quarks inside the proton, one may relate
${\cal W}^i$ to various quark distributions, in which
one avoids to use wave functions of the proton. Because
${\cal W}^i$ is defined with a twist-3 operator and
contributions from quark operators of twist-4 are presumably suppressed,
then these gluons are only generated by one quark. This can be illustrated
with Fig.2, where we use perturbative QCD as a guide and only three
gluon lines are drawn. In Fig.2. the black
dot represents the gluonic operator. Up to twist-3, all gluon lines
can only be attached to a single quark line, if gluon lines are attached
different quark lines, it will results in contributions at twist-4 level or higher.
Then ${\cal W}^i$ will be proportional to the quark density matrix:
\begin{equation}
 \langle p(p+\Delta)\vert \bar q_i (\frac{1}{2}\tau n ) q_j(-\frac{1}{2}\tau n)
 \vert p(p)\rangle,
\end{equation}
where $i,j$ stand for Dirac indices. If the calculation can be done
in this way with perturbative QCD, then one will find that ${\cal W}^i$
are related to various nondiagonal quark distributions, which may be found
in \cite{Diehl,HoodJi,Ji}. It is interesting to note that the helicity flip
quark distributions will contribute to $f_1$
in this calculation.
\par
It is constructive to look at how the differential cross-section of
$J/\Psi$ depends on
dimensional quantities to compare with that of $\eta_c$ in Eq.(50).
From \cite{MaXu} we have for $t\to 0$:
\begin{equation}
\frac{d\sigma}{dt}(J/\psi) \sim \alpha_s^2 \frac
{\vert \langle J/\psi \vert \psi^\dagger\bfsig\cdot\bfej({\bf v}=0)
      \chi \vert 0\rangle \vert^2}
{m_c^3}
 \cdot \frac{1}{s^2} \cdot \vert g_h(x_c)\vert^2,
\end{equation}
in this case we know that the dimensionless function corresponding
to $\hat f$ in Eq.(50) is determined by the usual gluon distribution
with $x_c$ defined after Eq.(47). For $x_c \to 0$ the gluon
distribution behaves like $x_c^{-(1+\beta)}$ with $\beta >0$,
this results in that the cross-section for $J/\Psi$ increases with
increasing $s$.
If we assume that $\hat f$ has a similar dependence on $s$ as
$g_h(x_c)$, we can roughly obtain the estimation:
\begin{equation}
\frac{\frac{d\sigma}{dt}(\eta_c)}{\frac{d\sigma}{dt}(J/\psi)}
\sim \alpha_s \frac{\vert \langle \eta_c \vert
\psi^\dagger \chi \vert 0\rangle\vert^2}
{\vert \langle J/\psi \vert \psi^\dagger\bfsig\cdot\bfej({\bf v}=0)
      \chi \vert 0\rangle \vert^2}\cdot \frac{\Lambda^2}{m_c^2}
      =0.008
\end{equation}
where the NRQCD matrix elements are determined by the decay
$\eta_c\to\gamma\gamma$ and $J/\Psi\to e^+e^-$, respectively,
the small scale $\Lambda$ is taken as $300$MeV, other parameters
are taken as $\alpha_s=\alpha_s(m_c)\approx 0.3$ and
$m_c=1.5$GeV. The estimation is rough
but it shows that the production $\eta_c$ is suppressed in comparison
with $J/\Psi$. Using the experimental data from HERA for $J/\psi$\cite{Herajpsi}
we estimate for $t\to 0$ or $\sqrt{\vert t\vert }\sim \Lambda=300$MeV:
\begin{equation}
  \frac{d\sigma}{dt}(\eta_c)\sim 2 {\rm nb/GeV}^2
\end{equation}
at $\sqrt{s}=100$GeV. This rough estimation gives a larger number than
those in \cite{engel1,engel2,Bartels}. For example, in \cite{engel2}
$d\sigma/dt$ is about $0.1{\rm nb/GeV}^2$ for $\sqrt{\vert t \vert}\sim 300$MeV.
In \cite{Dosch} it is pointed out that the proton-Odderon impact factor
used in \cite{engel1,engel2} may be overestimated, this will result in
that the prediction for $d\sigma/dt$ in \cite{engel1,engel2} may also
be overestimated. This makes that the difference between our estimation
and that in \cite{engel1,engel2} becomes larger than mentioned above.
It should be noted that  our results in Eq.(43) and Eq.(47)
can have large uncertainties.
Because $c$-quark may not be heavy enough,
corrections, like relativistic correction and those from higher orders in
$m_c^{-1}$, can be significant, as already discussed the case for $J/\Psi$.
For bottomonia, those corrections are expected to be small because
of the large $b$-quark mass.
\par
In our work we show that at the leading order of $m_c^{-1}$
the dominant contribution is from four distribution functions defined
with gluonic operator at twist 3. At higher orders operators at higher twist
will appear, because there are some higher-twist effect in the
3-gluon change and more than 3 gluons can be exchanged. One can
use some resummation techniques for gluon exchanges instead of
the expansion $m_c^{-1}$, like by using BFKL equation. In our case
with Odderon the equation for the resummation is
the Bartels-Kwiecinski-Praszalowicz (BKP) equation\cite{BKP}.
When the resummation is used,
the effect of higher twist contributions in the 3-gluon exchange,
including that from twist 3
contribution,
will be taken into account. At moment, no information is available
for the matrix element with the twist-3 operator in Eq.(48),
hence a detailed  prediction is not possible.
\par\vskip20pt

\noindent
{\large\bf 6. Contribution from Photon Exchange}
\par\vskip20pt
In this section we study the contribution from photon exchange instead
of exchange of many gluons in Fig.1. The contribution is represented
by Fig.3. We will only concentrate on the differential cross-section
in the extremely forward direction, i.e., in the limit $t\to 0$. In this limit,
the contribution is divergent at first look because the exchanged photon
becomes soft. But, it should be noted that for large and finite $s$,
$t$ can never be zero exactly, for large $s$, i.e.,
$s\gg M^2_{\eta_c}\approx (2m_c)^2$, $t$ approaches to
$-\vert t\vert_{min}$ in the nearly forward direction:
\begin{equation}
t\sim -\vert t\vert_{min},\ \ \ \ \
\vert t\vert_{min} \approx \frac {m^2 (2m_c)^2}{s},
\end{equation}
this minimum value will acts as an infrared cutoff and makes the
$S$-matrix element finite in the forward direction. Exchange of soft
gluons can happen in combination with the photon exchange in Fig.3,
it will lead to contributions in the forward region
which are suppressed by $\Lambda^2/m_c^2$ and these contributions
can be neglected.
\par

\begin{figure}[hbt]
\centering
\includegraphics[width=8cm]{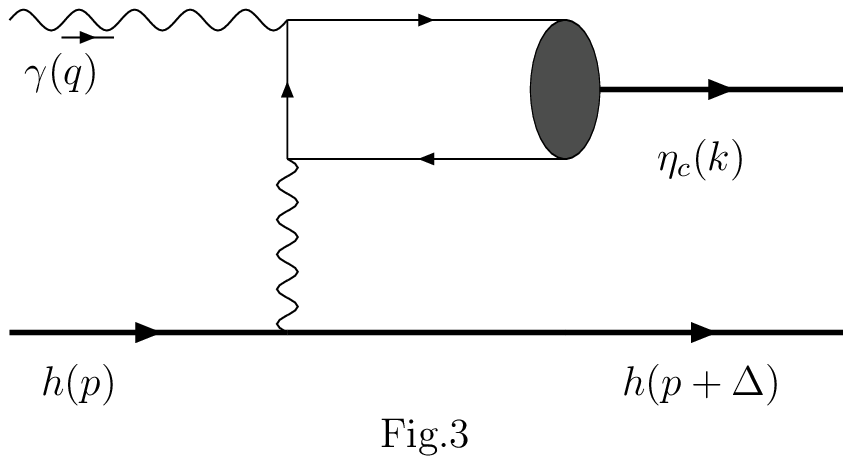}
\caption{One of the two diagrams for photon exchange}
\end{figure}
\par

The calculation of the $S$-matrix element is straightforward, we obtain:
\begin{eqnarray}
\langle f \vert S\vert i\rangle_{em} &=& i2(eQ_c)^2 (2\pi)^4 \delta^4(q-\Delta -k)
\frac{1}{\sqrt{v^0}}
\langle \eta_c \vert \psi^\dagger \chi \vert 0\rangle
  \varepsilon_{\mu\nu\sigma\rho} \varepsilon^\nu\Delta^\sigma v^\rho
  \nonumber\\
  && \cdot \frac{1}{\Delta^2+2m_c v\cdot\Delta}\cdot \frac{1}{\Delta^2}
     \cdot \langle h(p+\Delta)\vert J^\mu (0) \vert h (p)\rangle,
\end{eqnarray}
where $J^\mu$ is the operator of electric current, $Q_c$ is the charge
fraction of $c$-quark in unit $e$. For $t\to -\vert t\vert_{min}$,
one may use the method developed here
to define a wave function for the $c$- or $\bar c$ quark propagating
in a background field of soft photon to calculate the
$S$-matrix element, the same result can be obtained if one
neglects $\Delta^2$ in comparison to $2m_c v\cdot\Delta$ in the above equation.
In this way one can also show that the contribution from a photon exchange combined
with exchanging soft gluons is suppressed by $\Lambda^2/m_c^2$.
For $h$ being a proton, the matrix element
is decomposed with Dirac- and Pauli form factor:
\begin{equation}
\langle h(p+\Delta)\vert J^\mu (0) \vert h (p)\rangle
  = \bar u(p+\Delta ) \left \{ \gamma^\mu F_1(\Delta^2)
         +i\sigma^{\mu\nu} \Delta_\nu \frac{F_2(\Delta^2)}{2m}
         \right\} u(p),
\end{equation}
for $t=-\vert t\vert_{min}$ the differential cross-section can be obtained
directly. The dominant contribution reads:
\begin{eqnarray}
\frac{d\sigma}{dt}\vert_{t=-\vert t\vert_{min}}
   &\approx& \frac{\alpha_{em}^2Q_c^4}{8m_c^4}\cdot \frac{\vert
   \langle \eta_c \vert \psi^\dagger \chi \vert 0\rangle\vert^2}{m_c^3}
   \cdot \frac{s}{m^2}\cdot F_1^2(-\vert t\vert_{min})
   \nonumber\\
   &\approx &
   \frac{\alpha_{em}}{4m_c^5}\cdot \Gamma(\eta_c\to\gamma\gamma)
   \cdot \frac{s}{m^2},
\end{eqnarray}
where we use $F_1\approx e$ for the small momentum transfer. For large
$t$, because the form factor $F_1(t)$ falls like $t^{-2}$\cite{LB} and
the form factor $F_2(t)$ falls like $t^{-3}$\cite{BJY},
the differential cross-section
will fall like $t^{-8}$. A model calculation in \cite{engel1} for
gluon exchange predicts that the differential cross-section falls $t^{-2}$.
Hence, at large $t$ the differential cross-section with the photon
exchange
is highly suppressed. Because the detailed information of these form factors
are not available, numerical result of this contribution to the
total cross-section can not predicted.
For small $t$, the differential cross-section increases as
$1/t$, at $t=-\vert t\vert_{min}$ it increases with $s$
linearly, this is because the exchange photon becomes soft and its invariant
mass decreases with $s^{-1}$. From Eq.(58) the differential cross-section
in the forward region can be large, although it is suppressed
by $\alpha_{em}^3$. The numerical value of $d\sigma/dt$ can be worked out,
we have:
\begin{equation}
\frac{d\sigma}{dt}_{t=-\vert t\vert_{min}}\approx 8.4\times 10^{-4}\left( \frac{\sqrt{s}}{1{\rm GeV}}\right)^2
 \frac{{\rm nb}}{\rm GeV^2}.
\end{equation}
where we used $m_c=1.5$GeV.
With this estimation one can see that at $\sqrt{s}=100$GeV,
the contribution from photon exchange is already at the same
order as
the estimation given in Eq.(54). It is also larger
than the numerical values given in \cite{engel1,engel2,Bartels}.
Hence, if the production of $\eta_c$ is observed in the extremely forward
region the contribution from photon exchange is dominant for the production
at energies relevant to HERA.  This brings the difficulty to identify
the effect due to Odderon exchange in the extremely forward region. For $s$ larger
than those at HERA the contribution will still be dominant, if the contribution
from Odderon exchange increases with $s$ as $s^\alpha$ for $\alpha<1$. Since
the contribution from photon exchange decreases with increasing $\vert t\vert$
more rapidly than that of Odderon exchange, it is possible to identify
the effect of Odderon exchange for suitable $\vert t\vert $ which
is not very close to $\vert t\vert_{min} $. It requires a detailed study to
identify kinematical regions for hunting Odderon.

\par\vskip20pt
\noindent
{\large\bf 7. Summary}
\par\vskip20pt
In this work we have studied diffractive photoproduction of $\eta_c$.
Taking charm quark as a heavy quark, the nonpertubative effect
related to $\eta_c$ is represented by a NRQCD matrix element and
$\eta_c$ can be taken as a bound state of $c\bar c$ quark. Then
the production can be imagined as that the initial photon
splits into a $c\bar c$ pair, this pair forms the $\eta_c$ after exchange
of soft gluons with the initial hadron in the forward region.
The problem of exchange of soft gluons can be studied with HQET and a systematic
expansion in $m_c^{-1}$ can be employed. We find that the $S$-matrix element
can be expressed in terms of
wave-functions of $c$- and $\bar c$ quark,
which propagates under a background field of gluons. The background field
is dominated by components with long-wave lengthes corresponding to that
the exchanged gluons are soft. By solving these wave functions with the expansion
in $m_c^{-1}$, we can express the result for the $S$-matrix element
with quantities, which are well defined in QCD. The $S$-matrix element
in the forward region is expressed in the case of proton
with four functions, which are defined with a twist-3 operator of gluons.
If one thinks that the exchange of Odderon is responsible for the production,
then the effect of Odderon is represented by these functions.
For completeness, we also derive the $S$-matrix element for diffractive
photoproduction of $J/\Psi$, the result is exactly the same as that derived
in \cite{MaXu} with an assumption.
\par
Since these four functions are unknown, numerical predictions can not be made
in this work. However, an estimation in order of magnitude can be given
in comparison with the production of $J/\Psi$. The estimation indicates
that the differential cross-section of $\eta_c$ in the forward region
is really small, in qualitative agreement
with previous studies\cite{engel1,engel2,Bartels},
where some models are used to make numerical predictions. However, from our
results the differential cross-section of $\eta_c$ is not zero in the limit
$t\to 0$, this is different than those predicted in
\cite{engel1,engel2,Bartels}. The reason for this nonzero is that the initial
hadron can change its helicity in this limit.
\par
Instead of exchanging many soft gluons, a photon exchange can also contributes
to the production of $\eta_c$. In the forward direction, i.e., $t\to 0$,
one may find
that this contribution is divergent. But, we should note that $t$ can approach
to $0$ for $s \to \infty$, for large and finite $s$, the minimum value
of $\vert t\vert$ is never zero because all particle in the final state
are with finite masses. This minimum value makes the contribution finite.
At this minimum value of $\vert t\vert$ the differential cross-section due
to photon exchange can be predicted. We find that this contribution is large
and is dominant for production of $\eta_c$. Hence, if $\eta_c$ is produced
in the extremely forward region in experiment, one can not conclude that
the $\eta_c$ is produced through Odderon exchange and the effect of Odderon
is observed. At large $\vert t\vert $, it can be shown that the differential
cross-section with the photon exchange falls like $t^{-8}$ and can be neglected.
Therefore,
if $\eta_c$ is produced with $\vert t \vert$ which is not very close
to its minimal value, one can identify the production as an
effect of Odderon to confirm the existence of Odderon in this process.
\par
Our results for scattering amplitudes are expressed with quantities like
NRQCD matrix element and matrix element of the twist-3 operator of gluons,
they are well defined in the framework of QCD and are universal, i.e., they
do not depend on specific processes.
Nonperturbative methods like lattice QCD or sum rule method
may be used to determine them. The matrix element of the twist-3
operator is unfortunately unknown at the moment,
this fact prevents us to make numerical predictions.
But, with our results one can build more realistic models to have
a reliable prediction for diffractive photoproduction
of $\eta_c$.
\vskip 5mm
\begin{center}
{\bf\large Acknowledgments}
\end{center}
The author would like to thank Prof. O. Nachtmann and Prof. J.S. Xu
for interesting discussions. He would also like to thank Prof. X.D. Ji
for reading the manuscript and comments.
This work is supported by National Nature
Science Foundation of P. R. China with the grand No. 19925520.
\par\vskip20pt

\vfil\eject

\end{document}